# The Relativistic Bound States of a Non Central Potential

M. Eshghi [*,1], H. Mehraban [2], S. M. Ikhdair [3,4]

[1] *Yang Researchers and Elite Club, Central Tehran Branch, Islamic Azad University, Tehran, Iran*

[2] *Department of Physics, Semnan University, Semnan, Iran*

[3] *Department of Physics, Faculty of Science, an-Najah National University, Nablus, Palestine*

[4] *Department of Electrical Engineering, Near East University, Nicosia, Northern Cyprus, Mersin 10, Turkey*

## Abstract

We investigate the relativistic effects of a moving particle in the field of a pseudo-harmonic oscillatory ring-shaped potential under the spin and pseudo-spin symmetric Dirac wave equation. We obtain the bound state energy eigenvalue equation and the corresponding two-components spinor wave functions by using the formalism of suppersymmetric quantum mechanics (SUSYQM). Furthermore, the non-relativistic limits are obtained by simply making a proper replacement of parameters. The thermodynamic properties are shortly studied. Our numerical results for the energy eigenvalues are presented too.

## 1. Introduction

It is known that the study of the relativistic wave equations plays an important role in different fields of the modern physics. This started, by solving the spin-1/2 Dirac equation, as one of the most challenging wave equations in the past 80 years. The Dirac wave equation is mainly used in description of particles dynamics in the relativistic quantum mechanics, the behavior of nucleons in nuclei, and the relativistic collisions of heavy ions and recent interaction of laser with matter. Recently, many researchers have been working on the exact solution of the Dirac equation with different non-central potentials [1-5]. The near realization of these symmetries may explain degeneracies in some heavy meson spectra (spin symmetry) or in single-particle energy levels in nuclei (pseudo-spin symmetry) [6-8]. The concepts of spin and pseudo-spin symmetries are $SU(2)$ type symmetries of a Dirac Hamiltonian. They have been studied since 1969 in quasi-degeneracy. Besides, these symmetries were considered in the context of the deformed nuclei [9], the super-deformation [10], the magnetic moment interpretation [11, 12], the identical bands [13-16], and the effective shell-model coupling scheme [17]. The pseudo-spin symmetry occurs in the Dirac equation when $S(r) = -V(r)$, and the spin symmetry which is relevant to mesons appears in the Dirac equation when the subtraction of the scalar potential component $S(r)$ from time-like vector potential component $V(r)$ is equal to zero [18-20].

[*] ***Corresponding Author Email***: m.eshghi@semnan.ac.ir; eshgi54@gmail.com .

The non-central potentials are applied mainly in quantum chemistry and atomic physics. It is extensively used to describe many properties of some ring-shaped organic molecules (such as benzene molecular model) and also to study the interactions between deformed nucleons, that is, they are widely used in quantum chemistry and nuclear physics. Therefore, it is interested and necessary to study the solution of the Schrödinger, Klein-Gordon and Dirac wave equations with such non-central potentials [21-34]. The ring-shaped pseudo-harmonic oscillatory (RSPHO) potential is one kind of these physical potentials.

The spherical interaction potential takes the most general form:

$$V(r,\theta) = \frac{1}{2}Kr^2 + \frac{A}{r^2} + \frac{B}{r^2\sin^2\theta} + C\frac{\cos^2\theta}{r^2\sin^2\theta}, \qquad (1)$$

where $K, A, B$ and $C$ are constant parameters [35]. Figures 1 to 4 show plots of the RSPHO potential (1) for the given set of parameters values: $A = B = 0.01$, $C = 0.1$ and $K = 0.001$ as explained in each figure.

This paper is organized as follows. In Section 2, we solve the Dirac equation with the RSPHO potential (1) in the presence of the pseudo-spin and spin symmetries. We obtain the energy eigenvalue equations and the corresponding spinor wave functions by using the supersymmeric quantum mechanics (SUSYQM) method. In addition, we calculate some numerical results for the energy eigenvalues equation. Further, we find the non-relativistic limits of our solution by simply taking a proper replacement of parameters. We briefly discuss the thermodynamic properties of this non-central RSPHO potential. Finally, we present our conclusions in Section 3.

## 2. Solution of the Dirac Equation

The Dirac Hamiltonian in the natural units of $\hbar = c = 1$ is [36, 37]

$$H = \vec{\alpha} \cdot \vec{p} + \beta(M + S(\vec{r})) + V(\vec{r}), \tag{2}$$

where $S(\vec{r})$ and $V(\vec{r})$ stand for scalar and time-like vector non-central RSPHO potential, respectively, $\vec{\alpha}$ and $\beta$ are Dirac matrices, and $M$ denotes the composite fermionic mass. Thus, the Dirac equation can be written as

$$\left[\vec{\alpha} \cdot \vec{p} + \beta(M + S(\vec{r})) + V(\vec{r})\right] \Psi(\vec{r}) = E \Psi(\vec{r}), \tag{3}$$

where $E$ denotes the binding energy and the momentum $\vec{p} = -i\vec{\nabla}$. In the Pauli-Dirac representation, let us define the two spinor-components wavefunction:

$$\Psi(\vec{r}) = \begin{pmatrix} \varphi(\vec{r}) \\ \chi(\vec{r}) \end{pmatrix}, \tag{4}$$

so that we can obtain the following two coupled equations:

$$(\vec{\alpha} \cdot \vec{p}) \chi(\vec{r}) = [E - M - V(\vec{r}) - S(\vec{r})] \varphi(\vec{r}), \tag{5}$$

and

$$(\vec{\alpha} \cdot \vec{p}) \varphi(\vec{r}) = [E + M - V(\vec{r}) + S(\vec{r})] \chi(\vec{r}). \tag{6}$$

The spin symmetry demands that the scalar potential is to be equal to the time-like vector potential, that is, $S(\vec{r}) = V(\vec{r})$, so we have the following decoupled equations for the upper and lower spinor components of the wave function:

$$\left[p^2 + 2(E + M)V(\vec{r})\right] \varphi(\vec{r}) = (E^2 - M^2) \varphi(\vec{r}), \tag{7}$$

and

$$\chi(\vec{r}) = \frac{(\vec{\sigma} \cdot \vec{p})}{(E + M)} \varphi(\vec{r}), \tag{8}$$

respectively.

After substituting Eq. (1) into Eq. (7), we can obtain a second-order Schrödinger-like differential equation for the upper-spinor component as

$$\left\{-\nabla^2 + 2(E+M)\left[\frac{K}{2}r^2 + \frac{A}{r^2} + \frac{B}{r^2\sin^2\theta}\right.\right.$$
$$\left.\left. +C\frac{\cos^2\theta}{r^2\sin^2\theta}\right]\right\}\varphi(\vec{r}) = (E^2 - M^2)\varphi(\vec{r}). \tag{9}$$

Further, we need to make a separation of variables by inserting the following form of the upper component of the wave function given by

$$\varphi(\vec{r}) = \frac{R(r)}{r}G(\theta)F(\phi), \tag{10}$$

into Eq. (9) and this leads to the following set of second-order differential equations:

$$\left\{-\frac{d^2}{dx^2} + K(E+M)r^2 \right.$$
$$\left. +[2A(E+M)+\lambda]\frac{1}{r^2}\right\}R(r) = (E^2 - M^2)R(r), \tag{11}$$

$$\left\{-\frac{d^2}{d\theta^2} - \cot\theta\frac{d}{d\theta} - \lambda - \frac{d^2}{d\theta^2} - \cot\theta\frac{d}{d\theta} - \lambda \right.$$
$$\left. + \frac{2(E+M)(B+C\cos^2\theta)+m^2}{\sin^2\theta}\right\}G(\theta) = 0, \tag{12}$$

and

$$\frac{d^2F(\phi)}{d\phi^2} + m^2 F(\phi) = 0, \tag{13}$$

where $\lambda = \ell(\ell+1)$ and $m^2$ are two separation constants. The general solution to Eq. (13) is

$$F(\phi) = \frac{1}{\sqrt{2\pi}}e^{im\phi}, \quad m \in Z = 0, \pm 1, \pm 2, \dots . \tag{14}$$

### 2.1 *Solution of the angular part*

Now, we seek to find a solution for the angular part of wave function $G(\theta)$. Now letting

$$G(\theta) = H(\theta)/\sqrt{\sin\theta}, \tag{15}$$

and inserting it into Eq. (12), one can obtain the Schrodinger-like equation:

$$\left(\frac{d^2}{d\theta^2}+\tilde{V}\cot^2\theta\right)H(\theta)=\tilde{E}H(\theta),\ 0\leq\theta\leq\pi \tag{16}$$

where we have identified

$$\tilde{E}=-\lambda-\frac{1}{2}+2(E+M)B+m^2,\quad \tilde{V}=-2(E+M)(B+C)-m^2+\frac{1}{4}, \tag{17}$$

and with the requirements that the function $H(\theta=0)=H(\theta=\pi)=0$, that is, must vanish at the end points.

Now, we need to solve Eq. (16) by using the basic concepts of the SUSYQM formalism [38-41]. We can start out by writing down the ground-state lower spinor component $G_0(\theta)$ as

$$G_0(\theta)=\exp\left(-\int W(\theta)d\theta\right), \tag{18}$$

with $W(\theta)$ being called the superpotential in the SUSYQM formalism. Hence the substitution of Eq. (18) into Eq. (16) leads to the following equation satisfying $W(\theta)$ as

$$W^2(\theta)-W'(\theta)=\tilde{V}\cot^2\theta-\tilde{E}_0. \tag{19}$$

Taking the superpotential form as $W(\theta)=-Q\cot\theta$ and substituting it back into Eq. (19) we obtain the conditions,

$$Q=\tilde{E}_0, \tag{20}$$

and

$$Q^2-Q=\tilde{V}, \tag{21}$$

where

$$Q=\frac{1}{2}\pm\sqrt{\frac{1}{4}+\tilde{V}}. \tag{22}$$

Thus, the SUSYQM partner potentials $V_+(\theta)$ and $V_-(\theta)$ are given by

$$V_+(\theta)=W^2(\theta)+W'(\theta) \\ =(Q^2+Q)\cosec^2\theta-Q^2=(Q^2+Q)\cot^2\theta-Q, \tag{23}$$

and

$$V_-(\theta) = W^2(\theta) - W'(\theta) \qquad (24)$$
$$= (Q^2 - Q)\cosec^2\theta - Q^2 = (Q^2 - Q)\cot^2\theta - Q,$$

respectively. Equations (23) and (24) demonstrate that $V_+(\theta)$ and $V_-(\theta)$ are varied in similar shapes.

If the condition $V_+(\theta, a_0) = V_-(\theta, a_1) + R(a_1)$ is to be satisfied, the partner Hamiltonians are called shaped-invariant in the jargon, where $a_1$ is a new set of parameters uniquely determined from the old set $a_0$ via the mapping $F: a_0 \mapsto a_1 = F(a_0)$ and $R(a_1)$ does not include the independent variable $\theta$. In such a case $E_n = \sum_{k=1}^{n} R(a_k)$.

Now, we can write out $V_+(\theta)$ as

$$V_+(\theta) = W^2(\theta) + W'(\theta) = (Q+1)[(Q+1)-1]\cosec^2\theta - (Q+1)^2 + (Q+1)^2 - Q^2 \qquad (25)$$

So the potential $V_-(\theta)$ is a shape invariant potential as defined with

$$a_0 = Q, \qquad (26)$$

and

$$a_1 = F(a_0) = a_0 + 1 = Q + 1, \qquad (27)$$
$$a_n = a_0 + n = Q + n. \qquad (28)$$

Hence the partner potentials $V_+(\theta)$ and $V_-(\theta)$ satisfy the relationship $V_+(\theta, a_0) = V_-(\theta, a_1) + R(a_1)$, and the remainder $R(a_n)$ can be obtained from the relation as

$$R(a_1) = -a_0^2 + a_1^2, \qquad (29)$$
$$R(a_2) = -a_1^2 + a_2^2, \qquad (30)$$
$$R(a_3) = -a_2^2 + a_3^2, \qquad (31)$$

.
.

$$R(a_n) = -a_{n-1}^2 + a_n^2. \qquad (32)$$

For example, we can obtain from, Eq. (29), the above relation $R(a_1) = 1 + 2Q$ and so forth.

The ground-state energy of $V_-(\theta)$ is zero. For the partner potential $V_-(\theta)$, the energy spectrum is given by Ref. [42- 44]

$$\tilde{E}_n^{(-)} = \sum_{k=1}^{n} R(a_k) = R(a_1) + R(a_2) + ... + R(a_n) = -a_0^2 + a_n^2 = n^2 + 2nQ. \tag{33}$$

$$\tilde{E}_0^{(-)} = 0 \tag{34}$$

Hence, we can obtain the energy spectra as

$$\tilde{E} = \tilde{E}_n^{(-)} + \tilde{E}_0 = n^2 + 2nQ + Q = (n+Q)^2. \tag{35}$$

To check the accuracy of our results, we may set $B=0$ and $\alpha=1$ into Rosen-Morse (trigonometric) potential of Fig. 6 [45]. Our results of Eqs.(24), (26), (27) and (35) turn out to be identical to those ones obtained before in Ref. [45]. That is, the present results look exactly same as in [45].

To conclude, in using, Eq. (35), (22) and Eq. (17), we can obtain $\lambda$ as follows:

$$\lambda = \left[ n + \frac{1}{2} \pm \sqrt{\frac{1}{2} - 2(E+M)(B+C) - m^2} \right]^2 + 2(E+M)(B+C) + m^2 - \frac{1}{2}. \tag{36}$$

where $\lambda = \ell(\ell+1)$.

The significance of finding $\lambda$ in Eq. (36) is that it is the key factor in finding the energy levels of the system in terms of the orbital angular momentum and the potential parameters. The eigenvalue equation (36) will be essential in finding energy states when the radial part Schrodinger equation is solved in the next section.

## 2.2 Solution of the radial part

Now we seek to find a solution for the radial part of the wave function in Eq. (9) by identifying

$$\Delta = K(E+M), \tag{37}$$

$$\delta = 2A(E+M) + \lambda, \tag{38}$$

and

$$\tilde{E} = E^2 - M^2, \tag{39}$$

Then, the radial part equation can be rewritten in a more compact form:

$$\left[-\frac{d^2}{dr^2}+V_{eff}(r)\right]R(r)=\tilde{E}R(r), \qquad V_{eff}(r)=\frac{\delta}{r^2}+\Delta r^2. \tag{40}$$

Considering Eq. (40), one may introduce the following operators [39, 40]

$$\hat{A}=\frac{d}{dr}-W(r), \qquad \hat{A}^+=-\frac{d}{dr}-W(r), \tag{41}$$

where $W(r)$ is the radial superpotential. In the SUSYQM formalism [38-41], we can write down the radial part of the ground-state lower spinor component $R_0(r)$ as $R_0(r)=\exp\left(-\int W(r)dr\right)$ which is inserted into Eq. (40) to provide the Riccati equation

$$W^2(r)-W'(r)=V_{eff}(r)-\tilde{E}_0, \tag{42}$$

for which we assume the superpotential of the simple form

$$W(r)=\Delta r+\frac{\delta}{r}, \tag{43}$$

For a solution satisfying the Reccati equation, the following restrictions on the ansatz parameters

$$\Delta^2=K(E+M), \qquad \delta^2+\delta=2A(E+M)+\lambda, \qquad 2\delta\Delta-\Delta=-\tilde{E}_0, \tag{44}$$

are to be satisfied. After solving the set of equations in (44), the parameters $\delta$, $\Delta$ and $\tilde{E}_0$ are found to have the forms

$$\delta=-\frac{1}{2}\left(1+\sqrt{1+4(2A(E+M)+\lambda)}\right), \qquad \Delta=\sqrt{K(E+M)}, \qquad \tilde{E}_0=\Delta(1-2\delta). \tag{45}$$

Obviously, we have chosen the negative solution as the appropriate solution of the quadratic equation in $\delta$ so that we can get a positive physical energy states $\tilde{E}_0$.

In accordance we can now construct the two supersymmetric partner potentials as

$$V_+(r)=W^2(r)+W'(r)=\frac{\delta(\delta-1)}{r^2}+\Delta^2 r^2+2\delta\Delta+\Delta, \tag{46}$$

and $\hspace{10cm}$ (47)

$$V_-(r)=W^2(r)-W'(r)=\frac{\delta(\delta+1)}{r^2}+\Delta^2 r^2+2\delta\Delta-\Delta.$$

The above two partner potentials possess the following relationship

$$V_+(r,a_0) = V_-(r,a_1) + R(a_1), \tag{48}$$

where $a_0 = \delta$, $a_1 = f(a_0) = a_0 - 1 = \delta - 1$, the remainder can be followed with equation $R(a_1) = 4\Delta(a_0 - a_1)$. From Eq. (48), we know that the two partner potentials $V_+(r)$ and $V_+(r)$ are shape-invariant potentials in the sense of Ref. [42] and they have similar shapes. Using the shape invariance approach [42] to determine the energy spectra, the ground-state energy of $V_-(r)$ is zero, namely, $\tilde{E}_0^{(-)} = 0$. For the partner potential $V_-(r)$, the energy spectrum is given by [28, 42-44]

$$\begin{aligned}\tilde{E}_n^{(-)} &= \sum_{k=1}^{n} R(a_k) = R(a_1) + R(a_2) + ... + R(a_n) \\ &= 4\Delta(a_0 - a_1) + 4\Delta(a_1 - a_2) + 4\Delta(a_2 - a_3) + ... + 4\Delta(a_{n-1} - a_n), \\ &= 4\Delta[\delta - (\delta - n)] = 4n\Delta, \qquad n = 0,1,2,....\end{aligned} \tag{49}$$

This leads us to the expression

$$\begin{aligned}\tilde{E}_n &= \tilde{E}_0 + \tilde{E}_n^{(-)} = \Delta(1 - 2\delta) + 4n\Delta, \\ E^2 - M^2 &= \sqrt{K(E+M)}\left[2 + \sqrt{1 + 4(2A(E+M) + \lambda)}\right] + 4n\sqrt{K(E+M)}.\end{aligned} \tag{50}$$

Therefore, we can obtain the energy eigenvalue equation as

$$E^2 - M^2 = 2\sqrt{K(E+M)}\left[2n + 1 + \sqrt{\frac{1}{4} + 2A(E+M) + \lambda}\right], \tag{51}$$

with the quantum number $n = 0,1,2,...$.

It is worthy to note that when taking $\lambda = \ell(\ell+1)$, Eq. (51) reduces into Eq. (51) of Ref. [35] which was obtained before for the potential (1) using the standard associated Legendre differential equation. Therefore after making use of Eqs. (36) and (51), the energy states of the potential (1) can be easily found

$$E - M = 2\frac{\sqrt{K}}{\sqrt{E+M}}\left\{2n + 1 + \left\{\frac{1}{4} + 2A(E+M) + \right.\right. \\ \left.\left. + \left[n + \frac{1}{2} \pm \sqrt{\frac{1}{2} - 2(E+M)(B+C) - m^2}\right]^2 + 2(E+M)(B+C) + m^2 - \frac{1}{2}\right\}^{\frac{1}{2}}\right\}, \tag{52}$$

In Figures 5 to 8, we show the behavior of the spin symmetric energy eigenvalues versus the potential parameters $A$ and $K$ for various states of the quantum number $n = 1, 2, 3$.

Figure 5 shows the influence of the parameter $K$ on the energy spectrum $E_s$. It is obvious that the energy approximately increases with the increasing of the parameter $A$ for different increasing parameter values of $K = 5.0, 5.5, 6.0$ and when we take the state $n = 1$. We also see that the energy increases with the increasing of the quantum number $n = 1, 2, 3$ when the value $K = 5$.

In Fig. 6, we also show the influence of the parameter $B$ on the energy spectrum $E_n$. It is obvious that the energy increases versus the parameter $A$ with the decreasing azimuthal quantum number $m = 2, 0, -2$ and the energy decrease with the decreasing values of $B = -0.48, -0.49, -0.50$ for $m = 2$. We also see that the energy has pseuodo-linear decreases with the increasing of the azimuth quantum number $m = -2, 0, 2$. Furthermore, in Fig. 7, we show the influence of the parameter $A$ on the energy spectrum $E_n$. We see that the energy increases with the increasing of the value of parameter $K$ for increasing the values of $A = 5.0, 5.5, 6.0$ when $n = 1$. We also see the energy increases linearly with the quantum number $n = 1, 2, 3$ when $A=5$. Finally, in Fig. 8, we show the influence of the parameter $B$ on the energy levels $E_n$. We see that the energy decreases linearly with the increasing of parameter $K$ for different values of $B = -0.40, -0.45, -0.50$ when $m = 2$. We also notice that the energy increases with the decreasing of azimuthal quantum number $m = 0, 1, 2$ when $B = -0.50$.

We give some numerical results for spin symmetric energy state in Table 1.

Table 1 presents the calculated energy states with the changing of parameter $A$ while fixing the other parameters as $B = -0.05$, $K = 5$, $C = 0.005$ and $M = 5.0 \, fm^{-1}$.

At first the energy splitting increases with increasing $A$ and when we set $m = 0$, we find that the energy splitting increases with increasing $n$ values and it is becoming slightly smaller with increasing the value of $m$. In fact, the energy splitting decreases with increasing $m$.

We conclude that when $m = 0$, the angular part is related only to the change in parameter values of $B$ and $C$. Hence it has no much effect on our calculations to energy states. However, when $m > 0$, the energy states decrease by a smaller amount than the ones obtained when $m = 0$.

We have noticed a similar effect on the energy states while changing the parameter $A$. However, the essential feature is that the effect of this change is quite smaller than that made while changing the parameter $K$. This is due to that the parameter $K$. is the coefficient of harmonic oscillatory part while the parameter is $A$. the coefficient of pseudo-harmonic oscillatory part in RSPHO potential (1). That is the contribution of $K$. is larger than $A$ on the energy.

On the other hand to find a non-relativistic solution, we make the following simple mapping of parameters as $E - M \approx E_{NR}$, $E + M \approx 2\mu$, and in accordance the relativistic energy equation (52) and choosing $\xi = \mu(C + B)$, we have

$$E_{NR} = \frac{\sqrt{2K}}{\sqrt{\mu}} \left\{ 2n + 1 + \left\{ \frac{1}{4} + 4A\mu + + \left[ n + \frac{1}{2} \pm \sqrt{\frac{1}{2} - 4\xi - m^2} \right]^2 + 4\xi + m^2 - \frac{1}{2} \right\}^{\frac{1}{2}} \right\}, \quad (53)$$

On the other hand, let us investigate the thermodynamic properties, the energy has been calculated to obtain all thermodynamic quantities of the present non-relativistic model in a systematic way. So we should first calculate the partition function $Z$ at a finite temperature $T$, through the Boltzmann factor given by $Z = \sum_{n=0}^{\infty} e^{-\beta E_{n\ell}}$ where $\beta = 1/k_B T$ with $k_B$ is the Boltzmann constant. Hence the partition function $Z$ reads

$$Z = \sum_{n=0}^{\infty} \exp\left( -\frac{\beta\sqrt{2K}}{\sqrt{\mu}} \left\{ 2n + 1 + \left\{ \frac{1}{4} + 4A\mu + + \left[ n + \frac{1}{2} \pm \sqrt{\frac{1}{2} - 4\xi - m^2} \right]^2 + 4\xi + m^2 - \frac{1}{2} \right\}^{\frac{1}{2}} \right\} \right). \quad (54)$$

The other thermodynamic properties of the system can be easily found from the partition function. In fact, any other parameter that might contribute to the energy should also appear in the argument of Z [46]. Such as Helmholtz free energy $F$ which is alternatively defined as $F = -\ln(Z)^N/\beta$, the mean energy $U = -\partial \ln Z/\partial \beta$, the entropy is related to other quantities via $S = -\partial F/\partial T$ and the specific heat for higher temperatures can be found through $C = \partial U/\partial T$. Therefore, using the above equations, we can easily find the other thermodynamic quantities.

We can also find the wavefunctions of the radial part, in the spin symmetric case, through the relation [35]

$$R(r) = N e^{-\frac{\eta^2}{2}} \eta^{L+1} \times {}_1F_1\left(-n, L+\frac{3}{2}, \eta^2\right), \tag{55}$$

where N is the normalization constant, $\eta = r\sqrt[4]{(E+M)K}$ and $L(L+1) = 2(E+M)A + \lambda$.

The Dirac wave equation in the pseudo-spin symmetry when $S(r) = -V(r)$ takes the form

$$\left[p^2 + 2(E-M)V(r)\right]\varphi(\vec{r}) = \left[E^2 - M^2\right]\varphi(\vec{r}), \tag{56}$$

with the lower spinor component of the Dirac equation,

$$\varphi(\vec{r}) = \frac{(\vec{\sigma}.\vec{p})}{E-M} \chi(\vec{r}). \tag{57}$$

This pseudo-spin symmetry can be easily found by simply making the mapping transformations:

$$\phi(r) \to \chi(r), \quad \chi(r) \to -\phi(r), \quad V(r) \to -V(r), \quad E \to -E. \tag{58}$$

Hence the pseudo-spin symmetric radial wave functions can be obtained by substituting $\eta = r\sqrt[4]{(E-M)K}$ and $L(L+1) = 2(E-M)A + \lambda$ into Eq. (55). Further, the eigenvalue equation of the potential (1) can be obtained as:

$$E - M = 2\sqrt{\frac{-K}{E+M}} \left\{ 2n + 1 + \left\{ \frac{1}{4} - 2A(E+M) + \right. \right.$$

$$\left. \left. + \left[ n + \frac{1}{2} \pm \sqrt{\frac{1}{2} + 2(E+M)(B+C) - m^2} \right]^2 - 2(E+M)(B+C) + m^2 - \frac{1}{2} \right\}^{\frac{1}{2}} \right\}. \quad (59)$$

For the sake of completeness we plot Figs. 9 and 10 to show the relationship between the pseudo-spin energy states with the potential parameters and the two quantum numbers $n$ and $m$.

Further, we show the plot to the energy states against $B$ for different values of $A$ with $n$ and $K$ with $m$ in Figs. 11 and 12, respectively.

Also, in Table 2, we also calculate the energy states by changing the parameter $A$ while fixing the other parameters as $B = 0.50$, $K = -5.0$, $C=0.005$ and $M = 3.0\ fm^{-1}$. We see that, the energy splitting decreases with increasing $A$ and when we set $m = 0$, we find that the energy splitting increases with increasing $n$ values and it is becoming slightly smaller with increasing the value of $m$. In fact, the energy splitting decreases with increasing $m$.

A first look demonstrates the very approximate similar behavior with the spin symmetric case. The spacing between states becomes dense or far apart that is mainly dependent on the parameter values and the type of symmetry studied.

## 3. Discussions and Conclusions

In this work, we solved approximately the Dirac equation with spin and pseudo-spin symmetries for the PHORS potential (1) by means of the SUSYQM formalism. Approximate bound state energy eigenvalues and their associated two-component spinors of the Dirac particle are obtained in presence of the spin and psudo-spin symmetries. Our relativistic solution can be reduced to its non-relativistic limits once we make some appropriate mapping of parameters. Further, we also briefly discussed the thermodynamic properties of the resulting non-relativistic model.

Our numerical energy eigenvalues are obtained by taking some arbitrary numerical values of the parameters $K$ and $A$ and fixing the other parameters in the potential (1) for various principal and quantum numbers $n$ and $m,$ respectively. These results are displayed in Tables 1 and 2.

In the spin-symmetric energy states, it is noted that if $A$ and $n$ increase then $E$ increase, while if $m$ increases then $E$ decreases. However, in the pseudo-symmetric energy states if $A$ increases then $E$ decreases. If magnetic quantum number $m$ increases then $E$ decreases, while if principal quantum number $n$ increases then $E$ increases.

We have plotted the spin and pseudo-spin symmetries and shown the approximate similarity of energy in the presence of these two symmetries.

## Acknowledgment:

The authors would like to thank the kind referee(s) for positive and invaluable suggestions which have greatly improved the manuscript.

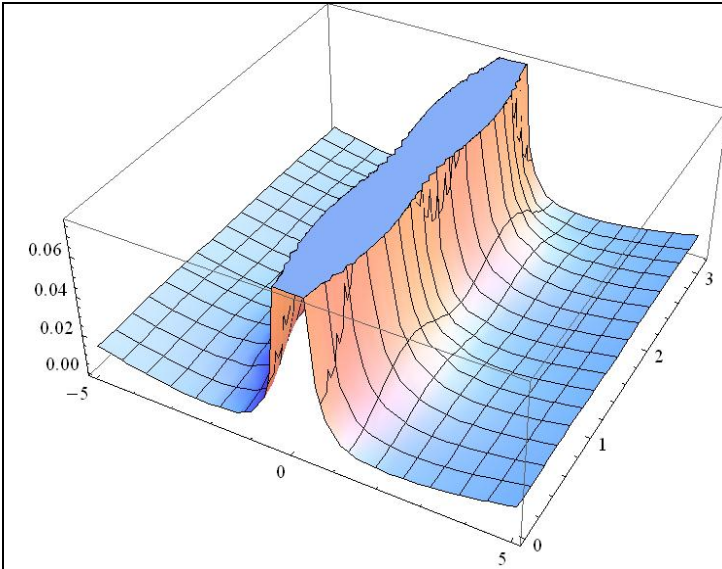

**Figure 1.** A plot of the RSPHO potential (1) in the range $r = [-5,5]$ and $\theta = [0, \pi]$ with values $A=B=0.01$, $C=0.01$ and $K=0.001$.

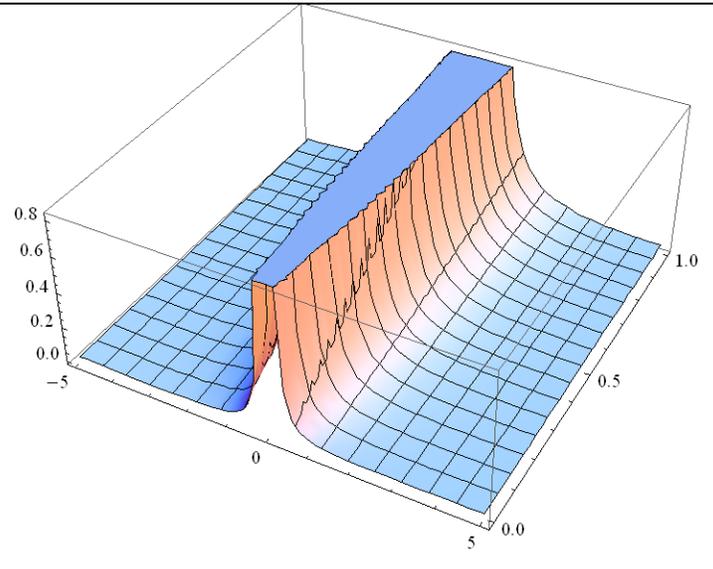

**Figure 2.** A plot of the RSPHO potential (1) in the range $r = [-5,5]$ and $B = [0,1]$ for values $\theta = \pi/4$, $A = 0.01$, $C=0.01$ and $K=0.001$.

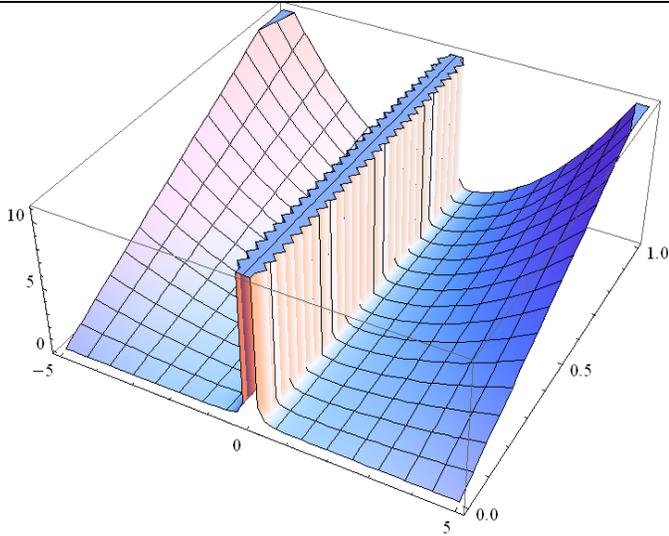

**Figure 3.** A plot of the RSPHO potential (1) in the range $r = [-5,5]$ and $K = [0,1]$ for sample $\theta = \pi/4$, $A=B=0.01$ and $C=0.01$.

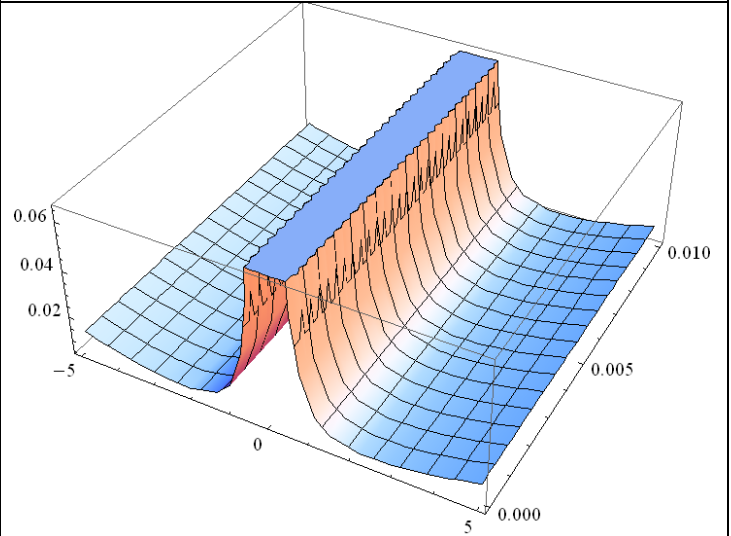

**Figure 4.** A plot of the RSPHO potential (1) in the range $r = [-5,5]$ and $C = [0,0.01]$ for values $\theta = \pi/4$, $A=B=0.01$ and $K=0.001$.

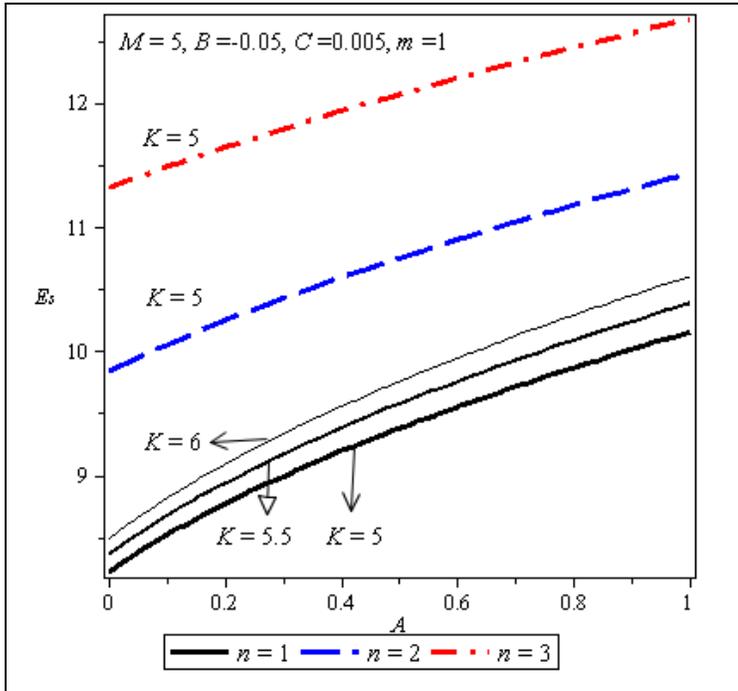

**Figure 5.** The spin symmetric energy states of the RSPHO potential versus $A$ for different $n$ and $K$

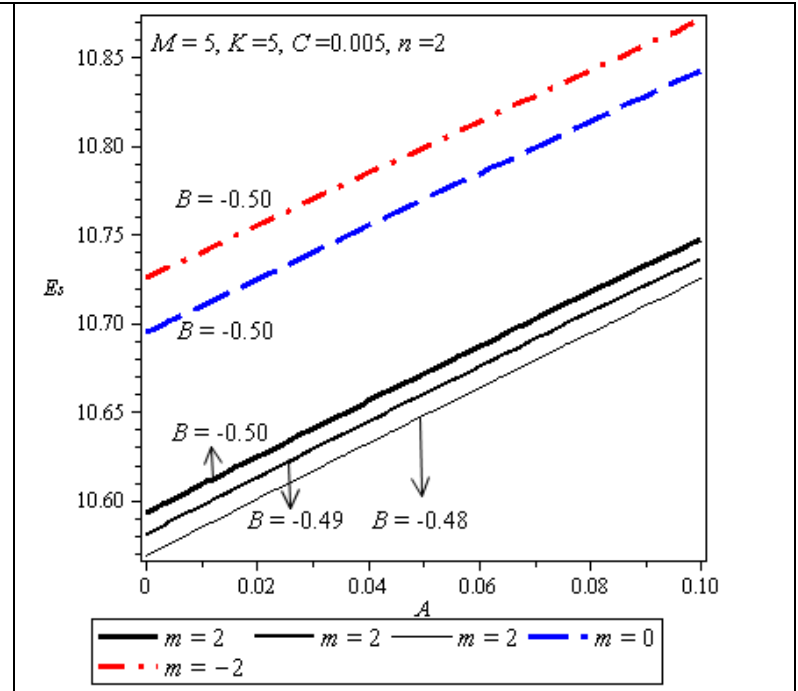

**Figure 6.** The spin symmetric energy states of the RSPHO potential versus $A$ for different of $m$ and $B$

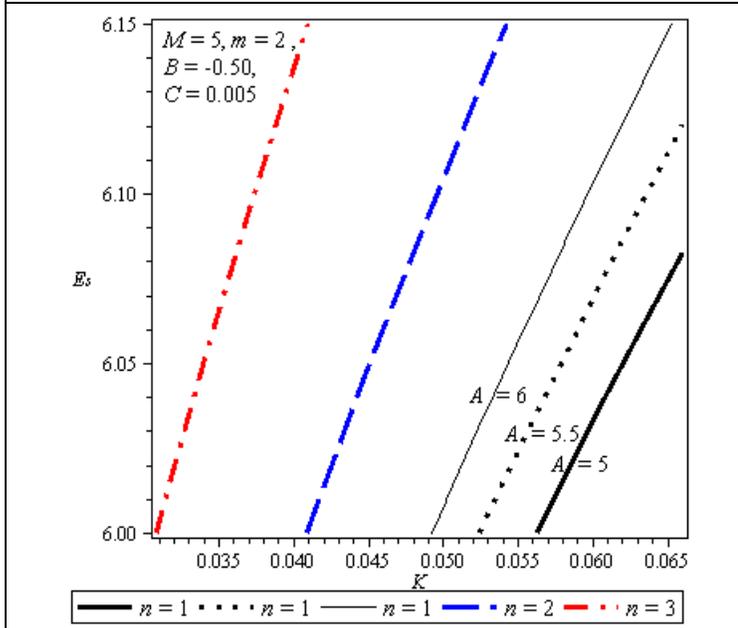

**Figure 7.** The spin symmetric energy states of the RSPHO potential versus $K$ for different $n$ and $A$

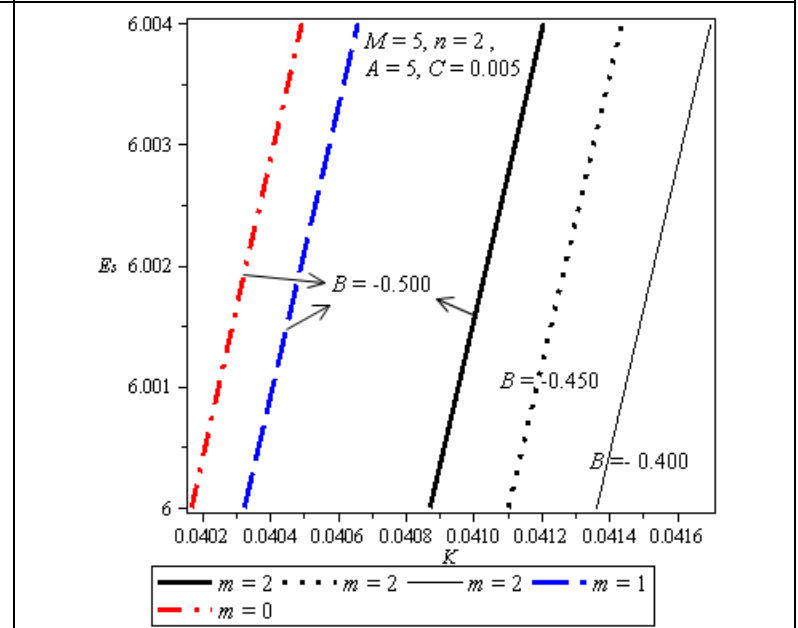

**Figure 8.** The spin symmetric energy states of the RSPHO potential versus $K$ for different $m$ and $B$

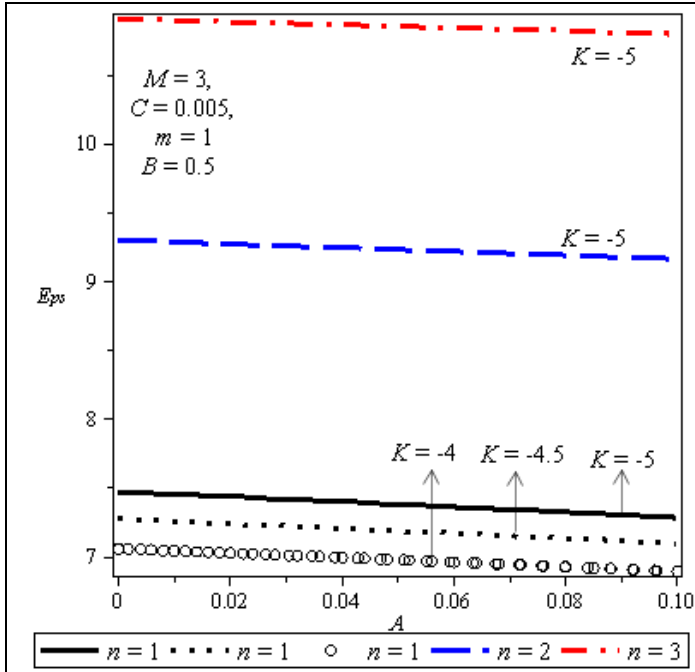

**Figure 9.** The pseudo-spin symmetric energy states of the RSPHO potential versus $A$ for different $n$ and $K$

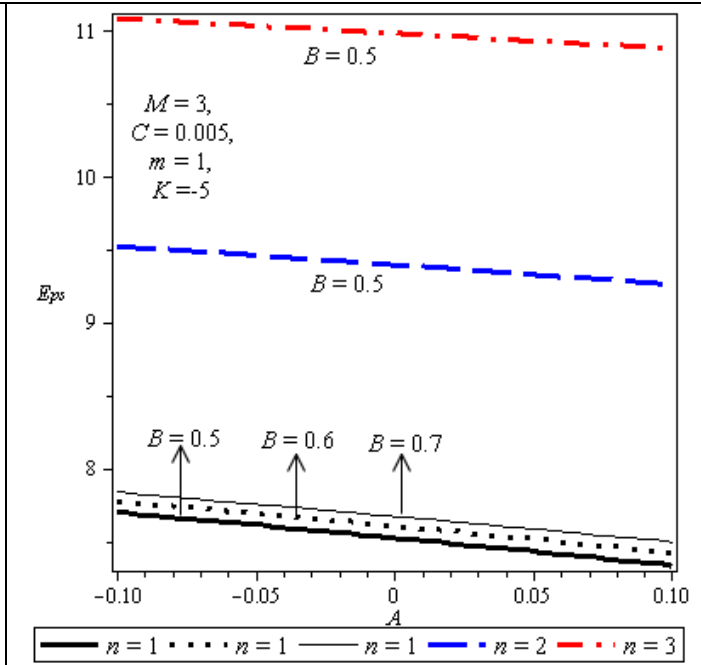

**Figure 10.** The pseudo-spin symmetric energy states of the RSPHO potential versus $A$ for different of $n$ and $B$

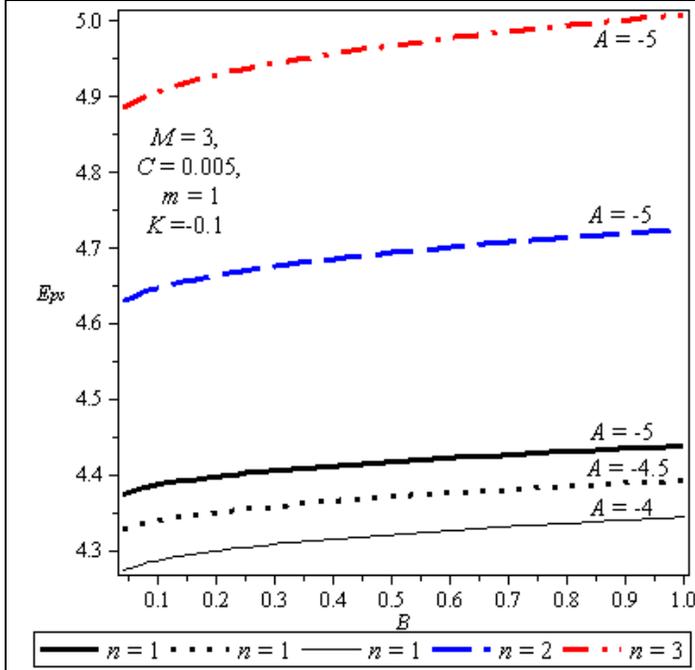

**Figure 11.** The pseudo-spin symmetric energy states of the RSPHO potential versus $B$ for different $n$ and $A$

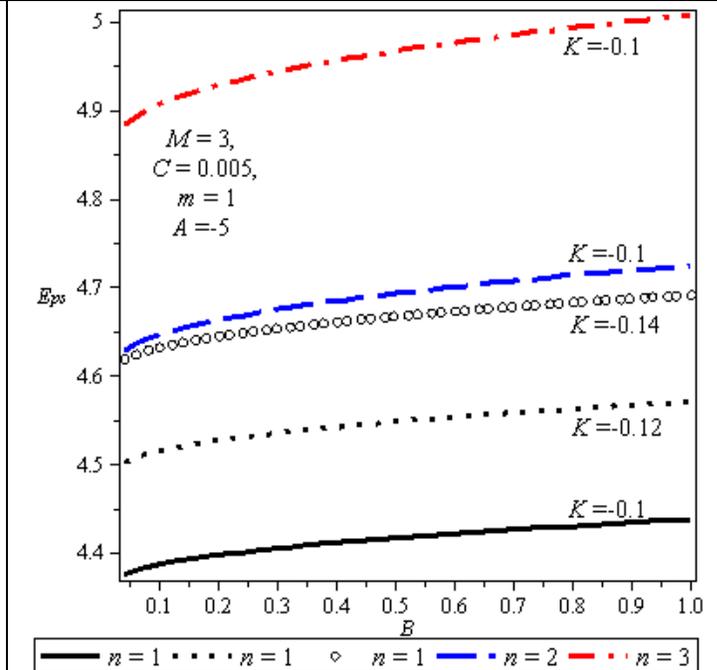

**Figure 12.** The pseudo-spin symmetric energy states of the RSPHO potential versus $B$ for different $m$ and $K$

Table. 1 The spin symmetric energy states for various values of A taking up B = -0.05, K = 5.0, C=0.005 and M =5.0 $fm^{-1}$

| n | m | A | $E_s$ | n | m | $E_s$ |
|---|---|---|---|---|---|---|
| 1 | 0 | 6 | 14.38516214 | 1 | 1 | 14.36707671 |
|   |   | 6.5 | 14.68410842 |   |   | 14.66709513 |
|   |   | 7 | 14.97241387 |   |   | 14.95634685 |
|   |   | 7.5 | 15.25115354 |   |   | 15.23592854 |
| 2 | 0 | 6 | 15.43922930 | 2 | 1 | 15.41239928 |
|   |   | 6.5 | 15.72755149 |   |   | 15.70222204 |
|   |   | 7 | 16.00603750 |   |   | 15.98203989 |
|   |   | 7.5 | 16.27564844 |   |   | 16.25284192 |
| 3 | 0 | 6 | 16.46852268 | 3 | 1 | 16.43488023 |
|   |   | 6.5 | 16.74677536 |   |   | 16.71490807 |
|   |   | 7 | 17.01595171 |   |   | 16.98566875 |
|   |   | 7.5 | 17.27690633 |   |   | 17.24804736 |

Table. 2 The pseudo-spin symmetric energy states for various values of A taking up B = 0.5, K = -5.0, C=0.005 and M =3.0 $fm^{-1}$

| n | m | A | $E_{ps}$ | n | m | $E_{ps}$ | n | m | $E_{ps}$ |
|---|---|---|---|---|---|---|---|---|---|
| 1 | 0 | -5 | 12.12523736 | 1 | 1 | 12.11721311 | 2 | 2 | 12.09120093 |
|   |   | -4.5 | 11.80243939 |   |   | 11.79377306 |   |   | 11.76561159 |
|   |   | -4 | 11.46422548 |   |   | 11.45480142 |   |   | 11.42409353 |
|   |   | -3.5 | 11.10808771 |   |   | 11.09775436 |   |   | 11.06397676 |
|   |   | -3 | 10.73074788 |   |   | 10.71930066 |   |   | 10.68174211 |
|   |   | -2.5 | 10.32777781 |   |   | 10.31492990 |   |   | 10.27258506 |
| 2 | 0 | -5 | 13.39533062 | 2 | 1 | 13.38420577 | 2 | 3 | 13.34829750 |
|   |   | -4.5 | 13.09302649 |   |   | 13.08111418 |   |   | 13.04258166 |
|   |   | -4 | 12.77772112 |   |   | 12.76489504 |   |   | 12.72330610 |
|   |   | -3.5 | 12.44751045 |   |   | 12.43360952 |   |   | 12.38840990 |
|   |   | -3 | 12.09996935 |   |   | 12.08478306 |   |   | 12.03524372 |
|   |   | -2.5 | 11.73192618 |   |   | 11.71517108 |   |   | 11.66030252 |
| 3 | 0 | -5 | 14.60737113 | 3 | 1 | 14.59415692 | 3 | 2 | 14.55167438 |
|   |   | -4.5 | 14.32247694 |   |   | 14.30842578 |   |   | 14.26316729 |
|   |   | -4 | 14.02646655 |   |   | 14.01145778 |   |   | 13.96301237 |
|   |   | -3.5 | 13.71786529 |   |   | 13.70174843 |   |   | 13.64960113 |
|   |   | -3 | 13.39483629 |   |   | 13.37741997 |   |   | 13.32091179 |
|   |   | -2.5 | 13.05504166 |   |   | 13.03607652 |   |   | 12.97434270 |